\documentclass[preprint,12pt]{elsarticle}

\usepackage{amssymb}
\usepackage{amsmath}
\usepackage{hyperref}
\usepackage{dsfont}
\usepackage{graphicx}
\usepackage{tikz}
\usepackage{multirow}
\usepackage{subcaption}
\usepackage{float}
\usepackage{graphicx}
\usepackage{mathtools}
\usepackage{microtype}
\usepackage{multicol}
\usepackage{geometry}
\usepackage{tabularx}
\usepackage{tikz}
\usepackage{url}
\allowdisplaybreaks

\newtheorem{theorem}{Theorem}

\newtheorem{definition}{Definition}
\usepackage{parskip}
\usetikzlibrary{arrows.meta,positioning,fit}

\journal{HAL}

\begin{document}

\begin{frontmatter}



\title{DFS-based fast crack pre-detection}

\author[label1]{Duc Nguyen\corref{cor1}}%
\ead{tran-1.nguyen@uni-ulm.de} 
\author[label2]{Vsevolod Chernyshev} 
\ead{vsevolod.chernyshev@uni-ulm.de} 
\author[label1]{Vitalii Makogin\corref{cor2}} 
\author[label1]{Evgeny Spodarev} 
\ead{evgeny.spodarev@uni-ulm.de} 

\cortext[cor2]{Dedicated to the memory of Dr. Vitalii Makogin (12.12.1987 - 08.05.2024)}
\cortext[cor1]{Corresponding author}

\affiliation[label1]{organization={Institute of Stochastics, Ulm University},
            addressline={Helmholtzstraße 16}, 
            city={Ulm},
            postcode={89081}, 
            state={Baden-Württemberg},
            country={Germany}}

\affiliation[label2]{organization={Institute of Optimization and Operations Research, Ulm University},
           addressline={Helmholtzstraße 16}, 
           city={Ulm},
           postcode={89081}, 
           state={Baden-Württemberg},
           country={Germany}}
\begin{abstract}
This paper develops a computationally efficient pre-detection method for cracks in three-dimensional CT images of concrete. Instead of attempting full voxel-wise crack segmentation, the method focuses on locating cubic subregions where crack structures are likely to be present and should be analyzed further. The proposed pipeline combines multiscale Maximal Hessian Entry filtering with graph-based connectivity analysis. After binarization, each subregion is represented by the boundary face with the largest foreground pixels, which transforms the local detection problem from a three-dimensional image task into a two-dimensional graph problem. A sparse lattice graph is constructed on the selected face, and Depth-First Search is applied to detect connected components corresponding to possible crack cross-sections. The choice of mesh size is justified by a probabilistic upper bound on a lattice-miss event. Experiments on semi-synthetic and real CT data show that the method gives fast, interpretable crack pre-localization while avoiding exhaustive analysis of the full image.
\end{abstract}




\end{frontmatter}



	\section{Introduction}
	
	Concrete is one of the principal construction materials used in buildings,
	bridges, and road infrastructure, which makes the reliable detection and
	analysis of cracks essential for structural safety, durability, and
	maintenance \cite{golewski}. To support such analyses, high-resolution
	X-ray computed tomography has become an important tool for investigating the internal microstructure of concrete, in which cracks
	typically appear as connected regions of low gray values. Therefore, a wide range of
	methods has been proposed to detect and segment these structures, ranging
	from classical multiscale filtering approaches
	\cite{westin2000, frangi98, ehrig2011comparison} to machine learning
	algorithms \cite{barisin2024riesz, Brox_2016, ilastik}, including recent
	deep learning models trained on semi-synthetic data
	\cite{dlcrack2024}. Comparative
	studies indicate that such methods can achieve strong performance
	\cite{BARISIN2022108747}, and the availability of annotated semi-synthetic data
	sets \cite{Jung2023, JUNG2024110474} has further supported their training
	and evaluation. However, for extremely large CT images of size, for
	example, $10000^2 \times 2000$ voxels, such as those produced by modern
	industrial CT systems \cite{gulliver2025}, the direct application of
	voxel-wise segmentation methods to the full volume becomes computationally
	prohibitive.
	
	A practical approach is to incorporate a pre-detection stage that identifies subregions likely to contain cracks and subsequently restricts precise segmentation to these localized areas. The goal of such a stage is not to produce a precise
	voxel-level crack map, but to reliably localize the subregions of the
	volume that contain crack structures, thereby limiting the domain for further analysis. Stochastic models for concrete constituents,
	such as Boolean models for pores \cite{Chiu2013, Baddeley2007} or random
	geometric surfaces for cracks \cite{Jung2023, JUNG2024110474}, offer one possible framework for pre-detection, but the complex and heterogeneous
	geometry of real CT data makes a model-free strategy more broadly
	applicable. In this paper, we therefore adopt a two-phase
	approach: a high-sensitivity Hessian-based binarization followed by a
	geometric examination of the subregion structure based on the Depth-First
	Search algorithm. The resulting framework prioritises computational efficiency and scalability, enabling rapid identification of crack-containing regions in large 3D CT images. Beyond the immediate
	computational savings from avoiding the processing of the vast crack-free
	majority of the volume, such pre-localization can also enhance the quality
	of any subsequent geometric or statistical analysis: by excluding background regions dominated by pores and aggregate boundaries, geometric statistics computed within the detected subregions more reliably characterize the crack structure.
	
	The choice of DFS is well motivated. Originating from the work of
	C.~Tr\'emaux and formalised by Tarjan \cite{tarjan1972depth} and Hopcroft
	and Tarjan \cite{hopcroft1973algorithm}, DFS identifies connected
	components in $O(\#V+\#E)$ time and has long been fundamental to
	connectivity analysis in binary images
	\cite{rosenfeld1966, rosenfeld1970, he2017ccl} and to graph connectivity
	problems \cite{even1975network}. Here, $\#V, \#E$ denotes the cardinality of $V$ and $E$, respectively. Its continued relevance in modern imaging
	applications is illustrated by several recent works: DFS-based
	post-processing has been used to connect fragmented crack predictions in
	infrastructure images \cite{mei2020dfscrack}, efficient
	connected-component labelling frameworks have been developed for large
	micro-CT volumes \cite{doube2021ccl}, and graph-based connected-component analysis has been applied to lesion tracking in CT oncology follow-up studies \cite{rochman2024graphct}. In the present setting, instead of traversing
	the full pixel graph, we construct a sparse boundary graph and apply DFS
	only there. This preserves the relevant crack geometry while keeping the
	computational cost low.
	\begin{figure}[htbp]
		\centering
		\begin{tikzpicture}[
			>={Stealth[length=3mm, width=2mm]},
			arrowstyle/.style={->, ultra thick, rounded corners}
			]
			
			\draw[dashed, rounded corners, fill=gray!10]
			(-0.5, -1.1) rectangle (10.5, 0.7);
			\node[anchor=south, font=\bfseries\small, color=black!80]
			at (5, -1.1) {Filtering};
			
			\draw[dashed, rounded corners, fill=gray!10]
			(-2.8, -3.6) rectangle (12.8, -1.6);
			\node[anchor=south, font=\bfseries\small, color=black!80]
			at (5, -3.6) {Graph-based Analysis};
			
			\draw[dashed, rounded corners, fill=gray!10]
			(-0.5, -6.1) rectangle (10.5, -4.1);
			\node[anchor=south, font=\bfseries\small, color=black!80]
			at (5, -6.1) {Classification};
			
			\node (step1) at (2.5, 0)
			[draw, rectangle, text width=3.8cm, align=center, fill=white, minimum height=1cm]
			{3D input image};
			\node (step2) at (7.5, 0)
			[draw, rectangle, text width=3.8cm, align=center, fill=white, minimum height=1cm]
			{Hessian Maximal Entry filter};
			
			\node (step3) at (-0.5, -2.5)
			[draw, rectangle, text width=3.8cm, align=center, fill=white, minimum height=1cm]
			{Image subdivision};
			\node (step4) at (5, -2.5)
			[draw, rectangle, text width=4.8cm, align=center, fill=white, minimum height=1cm]
			{Sparse subgraph construction};
			\node (step5) at (10.5, -2.5)
			[draw, rectangle, text width=3.8cm, align=center, fill=white, minimum height=1cm]
			{Connected components via DFS};
			
			\node (step6) at (5, -5.0)
			[draw, rectangle, text width=5cm, align=center, fill=white, minimum height=1cm]
			{Subregion classification \& crack pre-detection};
			
			
			\draw[arrowstyle] (step1) -- (step2);
			
			\draw[arrowstyle] (step2.south) -- ++(0,-0.85) -| (step3.north);
			
			\draw[arrowstyle] (step3) -- (step4);
			\draw[arrowstyle] (step4) -- (step5);
			
			\draw[arrowstyle] (step5.south) -- ++(0,-0.85) -| (step6.north);
			
		\end{tikzpicture}
		\caption{Processing pipeline for DFS-based crack pre-detection in 3D concrete CT images.}
		\label{fig:process_specified}
	\end{figure}
	
	The overall pipeline is summarised in
	Figure~\ref{fig:process_specified}. The first stage produces a binary
	candidate image with high sensitivity by means of the multiscale Maximal
	Hessian Entry filter. The second stage partitions this binary image into
	cubic subregions, selects the most informative face of each subregion,
	constructs a sparse graph, and applies DFS-based connected-component
	analysis to classify subregions as crack-containing or crack-free. In
	addition, the mesh size of the graph is chosen through an explicit
	probabilistic bound on the lattice-miss event for an idealised crack
	cross-section.
	
	The remainder of the paper is organised as follows.
	Section~\ref{filter} recalls the Maximal Hessian Entry filter used for
	high-sensitivity crack binarization. Section~\ref{sec:dfs} develops the
	DFS-based pre-detection procedure, including the reduction to
	two-dimensional face analysis, the lattice graph construction, the
	probabilistic mesh-size bound, and the connected-component classifier.
	Section~\ref{sec:experimental-evaluation} presents numerical experiments on
	both semi-synthetic and real CT images. Section~\ref{sec:discussion}
	discusses sensitivity and limitations, and
	Section~\ref{sec:conclusion} concludes the paper.
	
	\section{Crack segmentation}\label{filter}
	
	Before the DFS-based crack localization procedure (Section~\ref{sec:dfs}) can be applied,
	the raw 3D grayscale CT image must be converted into a binary image distinguishing crack
	voxels from background material. In this paper, we employ the Maximal Hessian Entry filter
	\cite{vitalii24}, a Hessian-based multiscale method well suited for detecting
	sheet-like, low-intensity structures in heterogeneous concrete. We briefly recall its
	definition here; further details and performance comparisons are given in \cite{vitalii24}.
	
	\textbf{Maximal Hessian Entry filter.}
	Let $I = \{I(p) \in [0,1],\ p \in W \subset \mathbb{Z}^3\}$ be a 3D grayscale image. For $\sigma > 0$, let $G(\,\cdot\,;\sigma)$ denote
	the 3-dimensional Gaussian kernel with independent coordinates and variance parameter $\sigma > 0$. The Hessian matrix of $I$
	at $p \in W$ is
	$$
	H(p;\,\sigma)
	=
	\Bigl(\sigma \cdot I(p) \ast
	\frac{\partial^2}{\partial p_i\,\partial p_j}\,G(p;\,\sigma)\Bigr)_{i,j=1}^{3},
	$$
	where $\ast$ denotes the convolution operation. Let $L_\sigma(I) = \{L_\sigma(I,p),\ p \in W\}$,
	where
	$$
	L_\sigma(I,\,p) = \max_{i,j=1,2,3}\bigl(H_{i,j}(p;\,\sigma),\;0\bigr),
	$$
	be the pointwise maximum over all six Hessian entries. For crack voxels lying on a
	nearly flat, low-intensity surface, at least one Hessian entry is expected to be strongly
	positive, reflecting the sharp intensity variation across the crack surface. Furthermore, replacing
	eigenvalue computation with the maximum-entry criterion significantly reduces
	computational cost while maintaining competitive detection accuracy
	\cite{BARISIN2022108747,vitalii24}.
	
	Denoting by $\mu(L_\sigma(I))$ and $\mathrm{sd}(L_\sigma(I))$ the sample mean and
	standard deviation of all values in $L_\sigma(I)$, the global threshold
	$$
	T_\sigma(I) = \mu(L_\sigma(I)) + 3\,\mathrm{sd}(L_\sigma(I))
	$$
	yields the single-scale binary image
	$$
	L_\sigma^*(I) = \{L_\sigma^*(I,p) = \mathbf{1}\{L_\sigma(I,p) \geq T_\sigma(I)\},\ p \in W\}.
	$$
	
	\textbf{Multiscale setting.}
	Because crack widths vary spatially across real CT images, a single scale $\sigma$ is insufficient. For a crack of constant width $w$, the recommended choice is $\sigma = w/2$. In practice, however, crack widths are neither constant nor known in advance. To address this, we consider a finite set of scales $\mathcal{S} \subset (0,\infty)$ and the Maximal Hessian Entry filter is defined as follows:
	$$
	L_\mathcal{S}(I) = \{L_\mathcal{S}(I,p) = \max_{\sigma \in \mathcal{S}}\,L_\sigma^*(I,p),\ p \in W\}.
	$$
	A voxel $p$ is classified as a potential crack voxel whenever it exceeds $T_\sigma(I)$ on at least one scale $\sigma \in \mathcal{S}$. In experiments, we use $\mathcal{S} = \{1,3,5,10\}$, covering crack widths from approximately 2 to 20 voxels. While this union strategy ensures high recall, $L_\mathcal{S}(I)$ inevitably contains false positives from air pores and aggregate boundaries, whose geometry differs from the elongated crack structure and can be eliminated by the DFS-based steps in Section~\ref{sec:dfs}.
	
\begin{figure}[H]
		\centering
		\includegraphics[width=0.9\textwidth]{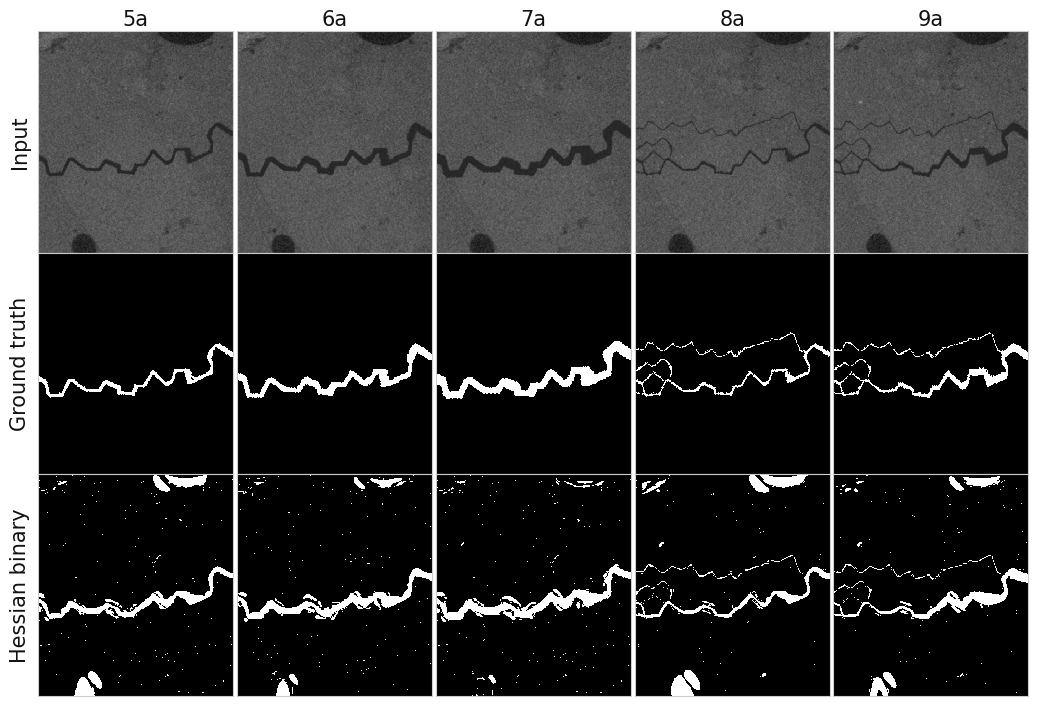}
		\caption{Maximal Hessian Entry filter ($\mathcal{S}=\{1,3,5,10\}$) applied to five
			semi-synthetic images from the NC subset of VoroCrack3d \cite{JUNG2024110474}.
		}
		\label{fig:hessian1}
	\end{figure}

\begin{figure}[H]
		\centering
		\includegraphics[width=0.75\textwidth]{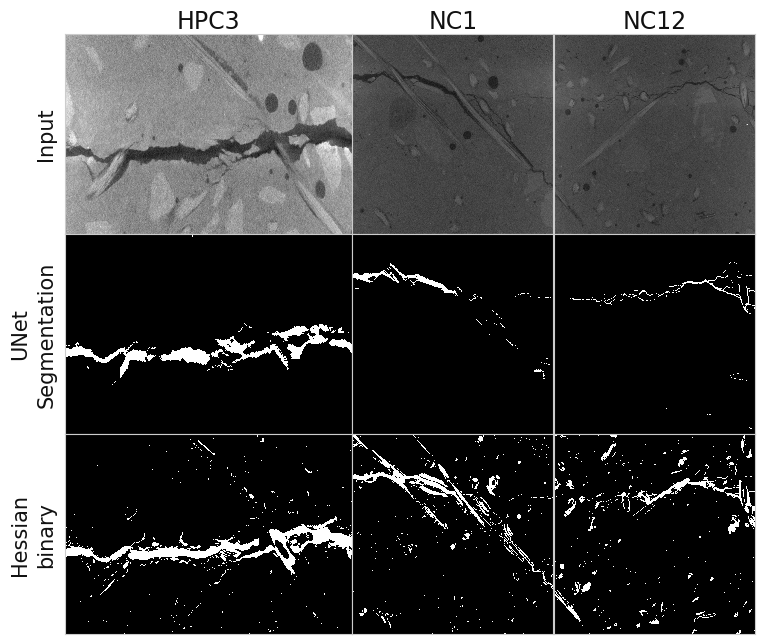}
		\caption{Maximal Hessian Entry filter ($\mathcal{S}=\{1,3,5,10\}$) on three real CT
			scans.
		}
		\label{fig:hessian2}
	\end{figure}

	\textbf{Performance.}
	Figures~\ref{fig:hessian1} and \ref{fig:hessian2} illustrate the filter output on two
	image collections. Figure~\ref{fig:hessian1} shows results on five semi-synthetic images
	from the Normal Concrete (NC) subset of VoroCrack3d \cite{JUNG2024110474}. For each image, the figure
	displays the input grayscale slice (top), the binary ground truth label (middle), and the
	binary output $L_\mathcal{S}(I)$ (bottom). The five images progress from fixed-width cracks of 3, 5, and 7
	voxels (images 5a, 6a, 7a) to multiscale branching cracks
	(images 8a, 9a). In all cases the filter recovers the crack 
	well, with false positives from air pores visible as small circular blobs in the binary
	output.
	Figure~\ref{fig:hessian2} shows results on three real CT scans: one sample of
	High-Performance Concrete (HPC3), a cement-based material with low water-to-binder ratio
	and high compressive strength, and two samples of Normal Concrete (NC1, NC12). Since no
	voxel-level ground truth is available for real images, the output of a U-Net \cite{dlcrack2024}
	trained on VoroCrack3d is used as a reference segmentation. For each scan, the figure
	displays the input slice (top), the U-Net segmentation (middle), and the Hessian binary
	output $L_\mathcal{S}(I)$ (bottom). The filter successfully recovers the main crack trajectory in all three cases, with a higher false positive rate than in the semi-synthetic images, reflecting the broader range of artefacts present in real CT data. 
	
	The results above demonstrate that the Maximal Hessian Entry filter reliably preserves
	the crack structure in the binary image $L_\mathcal{S}(I)$, which is the essential
	requirement for any subsequent processing. However, $L_\mathcal{S}(I)$ remains noisy,
	containing a substantial number of false positives from air pores and aggregate
	boundaries. The goal of the following section is to exploit the geometric elongation of
	cracks to localise crack-containing regions and suppress these false detections,
	thereby reducing the search space for subsequent exact crack segmentation methods.
		
	\section{DFS-based crack pre-detection}
	\label{sec:dfs}
	
	Recall that the objective of this section is not to obtain a detailed segmentation of
	cracks, but to localize subregions of the image that contain crack
	structures. To this end, a two-stage strategy is adopted: the binary image $L_\mathcal{S}(I)$ is first partitioned into cubic subregions, and each subregion is then classified as either crack-containing or crack-free. The classification procedure for each subregion is built as follows: a dimensional reduction from 3D to 2D boundary analysis, a graph-based representation of the binarized image reflecting the geometric structure of cracks, and a probabilistic bound that controls the rate of missed detections.
	
	\subsection{Reduction to two-dimensional surface analysis}
	\label{ssec:reduction}
	
	The binary image $L_\mathcal{S}(I)$ is partitioned into a regular grid of
	cubic subimages $\{A_q\}_{q \in \mathcal{Q}}$, where $\mathcal{Q}$ indexes
	the grid cells and each $A_q$ corresponds to a cubic subregion
	$\widetilde{W}_q = [a_q, b_q]^3 \subset \mathbb{R}^3$ of side length
	$n = b_q - a_q + 1$. Each subimage is examined independently to determine
	whether it contains a crack. Since cracks in concrete CT images are thin, elongated, and nearly planar, any sufficiently large crack in $\widetilde{W}$ must intersect its boundary $\partial\widetilde{W}$.
	Consequently, detecting a crack inside $\widetilde{W}$ reduces to
	detecting its presence on at least one of the six faces of
	$\partial\widetilde{W}$, which is a two-dimensional problem and
	considerably cheaper than processing the full volume.
	
\begin{figure}[H]
		\centering
		\includegraphics[width=\textwidth]{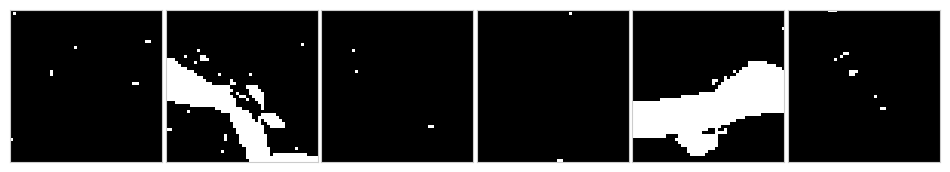}
		\caption{Most informative face of six randomly chosen subregions of size $50^3$.
			Crack-containing subregions exhibit a large connected foreground region on the
			selected face; crack-free subregions show only small spurious components.}
		\label{fig:facet2d}
	\end{figure}

	In practice, we select the face $\mathcal{W} \subset \partial\widetilde{W}$
	with the largest number of foreground pixels, since that face is most likely
	to exhibit a visible crack cross-section. Figure~\ref{fig:facet2d} shows
	the most informative face of six randomly chosen subregions of size $50^3$:
	in crack-containing subregions the crack appears as a large connected
	foreground region, while crack-free subregions show only scattered noise
	from air pores and aggregate boundaries.
	
	Formally, let $J = \{J(p) \in \{0,1\} : p \in \mathcal{W}\}$ denote the binary image
	restricted to the selected face, where $\mathcal{W} \subset \mathbb{Z}^2$,
	and foreground pixels ($J(p)=1$) correspond to the crack voxel as identified by the
	Maximal Hessian Entry filter. The crack detection procedure is then applied to $J$
	rather than to the full 3D subimage $A_q$.
	
	\subsection{Lattice graph construction}
	\label{ssec:graph}
	
\begin{figure}[htbp]
		\centering
		\includegraphics[width=\textwidth]{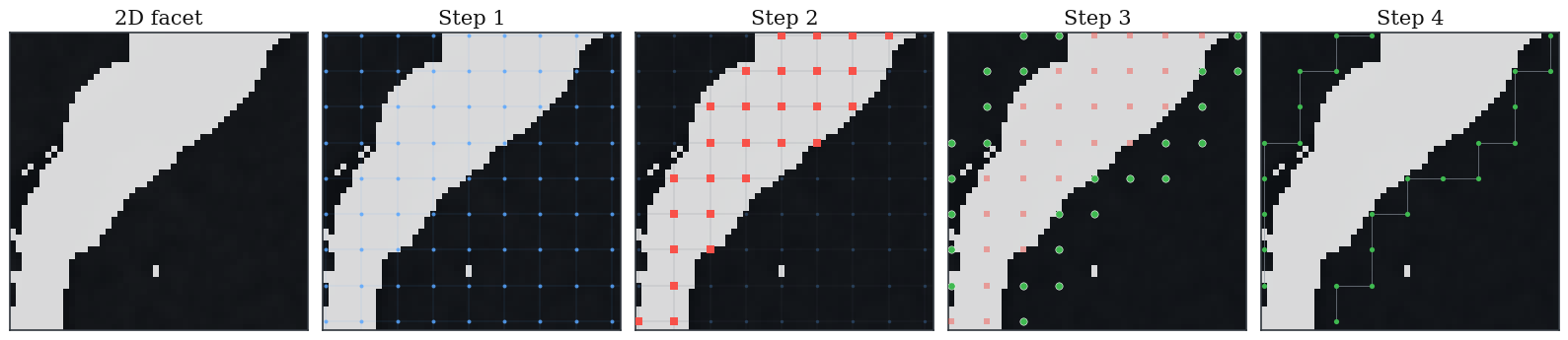}
		\caption{Construction of $G^* = (K, E_K)$ on a face of image 6a. From left to right:
			\textbf{1:} Binary facet $J$.
			\textbf{2:} Coarse lattice $G_\Delta$ (blue).
			\textbf{3:} Foreground nodes $H$ (red).
			\textbf{4:} Vertex boundary $K$ (green), $\ell^\infty$-adjacent to $H$ (pink).
			\textbf{5:} Graph $G^*$; connected components trace the crack boundary.}
		\label{fig:graph}
	\end{figure}

	Having reduced the problem to a 2D binary image $J$ on the selected face $\mathcal{W}$,
	we now describe how to efficiently detect the crack presence in $J$. A naive
	application of DFS to the full pixel graph of $J$ is possible but expensive: the vast
	majority of pixels in $J$ carry no crack-relevant information. We therefore construct
	a sparser graph $G^*$ that concentrates only on the neighbourhood of the foreground, 
	motivated by the observation from Section~\ref{ssec:reduction} that a crack
	forms a large connected foreground region on $\mathcal{W}$. The construction is illustrated in Figure~\ref{fig:graph}.
	
	\begin{definition}[Crack-sensitive lattice graph]
		\label{def:graph}
		Given a binary image $J$ on $\mathcal{W}$ and a mesh size $\Delta \in \mathbb{N}$,
		the lattice graph $G^* = (K, E_K)$ is constructed as follows.
		
		\begin{enumerate}
			\item \textbf{Lattice graph.}
			Define the grid $V_\Delta = \mathcal{W} \cap \Delta\mathbb{Z}^2$
			and the lattice graph $G_\Delta = (V_\Delta, E_\Delta)$, where
			$$E_\Delta = \bigl\{(e_1, e_2) \mid e_1, e_2 \in V_\Delta,\;
			\|e_1 - e_2\|_2 = \Delta\bigr\}.$$
			
			\item \textbf{Foreground identification.}
			Let $H = \bigl\{p \in V_\Delta : J(p) = 1\bigr\}$ be the subset of
			lattice nodes that coincide with foreground (crack) pixels.
			
			\item \textbf{Vertex boundary.}
			Define
			$$K = \bigl\{q \in V_\Delta \setminus H \mid
			\exists\, p \in H,\; \|q - p\|_\infty \leq \Delta\bigr\},$$
			the vertex boundary of $H$: the set of background lattice nodes
			whose $\ell^\infty$-neighbourhood intersects $H$. 
			
			\item \textbf{Induced subgraph.}
			Restrict the edge set to
			$E_K = \bigl\{(e_1, e_2) \in E_\Delta : e_1, e_2 \in K\bigr\}$
			and return $G^* = (K, E_K)$.
		\end{enumerate}

	\end{definition}
	
	The vertex boundary $K$ plays a central role in crack localization: its connected
	components in $G^*$ define the boundary of crack regions rather than their
	interior. As shown in Figure~\ref{fig:graph}, when a crack is present, $K$ forms
	elongated connected components that trace the crack boundary across the image.
	In the absence of a crack, $K$ is expected to consist of many small, nearly circular
	components enclosing individual air pores. The crack detection problem thus reduces
	to distinguishing these two cases, which is performed by the DFS-based classifier described in Section~\ref{ssec:classification}.
	
	The complexity of DFS on $G^*$ is $O(\#K + \#E_K)$, where $\#K$ denotes the
	cardinality of $K$. Since $\#K \ll \#\mathcal{W}$ for sufficiently large $\Delta$,
	this yields a substantial reduction in computational cost compared to operating 
	on the full pixel graph. However, $\Delta$ must be chosen carefully: if it is too large relative to the crack width
	$w_0$, the coarse grid $V_\Delta$ may miss the crack altogether, leaving $K$ empty
	or fragmented into small components that do not extend across $\mathcal{W}$, and hence
	indistinguishable from components generated by pores. Conversely, if $\Delta$ is too small, $G^*$
	approaches the full pixel graph and the computational gain is lost. The choice of $\Delta$ is therefore constrained by a bound on the misclassification probability, which we quantify in the following subsection.

	\subsection{Probability of missing a crack: theoretical bound}
	\label{ssec:probability}
	
	\textbf{Setup and error controlling.}
	Since the lattice $\Delta\mathbb{Z}^2$ is fixed and the crack position is unknown,
	the crack location is treated as random: let $\rho$ be a uniformly random
	isometry of $\mathbb{R}^2$, and define the miss probability as
	$$
	\mathbb{P}\bigl\{\mathcal{M}\bigr\}
	\;:=\;
	\mathbb{P}\bigl\{\rho(C_0) \cap \Delta\mathbb{Z}^2 = \emptyset\bigr\},
	$$
	where $C_0$ is a (for simplicity, convex) body in $\mathbb{R}^2$ representing the crack cross-section, with area $|C_0|$ and width $w_0$. By a standard scaling argument, the randomly placed body $\rho(C_0)$ missing $\Delta\mathbb{Z}^2$ is equivalent to the rescaled body
	$\widetilde{C}_0 = C_0/\Delta$ (with area $|\widetilde{C}_0| = |C_0|/\Delta^2$
	and width $\widetilde{w}_0 = w_0/\Delta$) missing $\mathbb{Z}^2$, so
	\begin{equation}\label{eq:scaling}
		\mathbb{P}\bigl\{\mathcal{M}\bigr\}
		\;=\;
		\mathbb{P}\bigl\{\rho'(\widetilde{C}_0) \cap \mathbb{Z}^2 = \emptyset\bigr\},
	\end{equation}
	where $\rho'$ is a uniformly random isometry of $\mathbb{R}^2$.
	It is worth noting that, if $\Delta \leq w_0$ then $\widetilde{w}_0 = w_0/\Delta \geq 1$, meaning the
	rescaled body $\widetilde{C}_0$ has width at least $1$, so every placement
	of $\widetilde{C}_0$ must contain a point of $\mathbb{Z}^2$. Hence
	$\mathbb{P}\{\mathcal{M}\} = 0$ and no further bound is needed.
	It therefore suffices to consider $\Delta > w_0$, equivalently $\widetilde{w}_0 < 1$. We apply the following result.

	\begin{theorem}[\protect{\cite[Theorem~4]{roldan2012probability}}]
		\label{thm:pensado}
		For every $\varepsilon > 0$ there exist constants $\bar{c}$ and $\bar{w}$
		such that if $C$ is a planar convex body with $|C| > \bar{c}$ and mean
		width less than $\bar{w}$, then
		$$\mathbb{P}\bigl\{\rho(C) \cap \mathbb{Z}^2 = \emptyset\bigr\}
		\;\leq\; \frac{1+\varepsilon}{4|C|}.$$
	\end{theorem}
	
	Applying Theorem~\ref{thm:pensado} to $\widetilde{C}_0$ and
	substituting back via~\eqref{eq:scaling} gives
	\begin{equation}\label{eq:bound_raw}
		\mathbb{P}\bigl\{\mathcal{M}\bigr\}
		\;\leq\;
		\frac{1+\varepsilon}{4|\widetilde{C}_0|}
		\;=\;
		\frac{\Delta^2(1+\varepsilon)}{4|C_0|},
	\end{equation}
	where $|\widetilde{C}_0| = |C_0|/\Delta^2$.
	
	\textbf{Dependence of $\varepsilon$ on parameters.}
	The parameter $\varepsilon$ in Theorem~\ref{thm:pensado} is not a free parameter
	but is determined by the geometry of $C_0$ and the ratio $w_0/\Delta$.
	The explicit bound derived in \cite[Section~5]{roldan2012probability} is
	\begin{equation}\label{eq:bound_full}
		\mathbb{P}\bigl\{\mathcal{M}\bigr\}
		\;\leq\;
		\frac{\Delta^2}{4|C_0|}
		\!\left(
		1
		+ C\frac{w_0}{\Delta}\log^2\!\!\left(\frac{w_0}{\Delta}\right)
		+ C\frac{\Delta^2}{|C_0|}
		\right),
	\end{equation}
	for some constant $C$.
	
	For simplicity, set
	\begin{equation}\label{eq:eps_shape}
		\varepsilon(\Delta)
		=
		\frac{w_0}{\Delta}\log^2\!\!\left(\frac{w_0}{\Delta}\right)
		+
		\frac{\Delta^2}{|C_0|},
	\end{equation}
	so that \eqref{eq:bound_full} becomes
	\begin{equation}\label{eq:bound_compact}
		\mathbb{P}\bigl\{\mathcal{M}\bigr\}
		\;\leq\;
		\frac{\Delta^2}{4|C_0|}
		\Bigl(1 + C\cdot\varepsilon(\Delta)\Bigr).
	\end{equation}
	Since $C$ is not known explicitly, we set $C = 1$ in all numerical computations.
	When $w_0/\Delta < 1$ and $|C_0|$ is large, both terms in
	$\varepsilon(\Delta)$ are small and the bound is dominated by
	$\Delta^2/(4|C_0|)$, so the constraint on $\Delta$ depends only weakly on the value of $C$.

	\textbf{Optimal mesh size selection.} A larger mesh size $\Delta$ reduces the cardinality of $V_\Delta$ and hence
	the computational cost of DFS on $G^*$, but increases the risk of missing
	the crack. The bound~\eqref{eq:bound_compact} quantifies this trade-off:
	given a tolerance $\alpha \in (0,1)$ on the miss probability, the condition
	$\mathbb{P}\{\mathcal{M}\} \leq \alpha$ is satisfied if
	\begin{equation}\label{eq:suff_cond}
		\frac{\Delta^2}{4|C_0|}
		\bigl(1 + \varepsilon(\Delta)\bigr)
		\leq \alpha.
	\end{equation}
	The left-hand side of~\eqref{eq:suff_cond} increases with $\Delta$ on $\Delta > w_0$. Whenever some integer $\Delta>w_0$ satisfies~\eqref{eq:suff_cond}, a largest such integer is well defined; if no integer exceeds the crack width while still meeting the tolerance, the requirement cannot be enforced and we keep the finest available lattice instead. This leads to the following definition.
	
	\begin{definition}[Optimal mesh size]
		\label{def:deltamax}
		Given $\alpha \in (0,1)$, crack width $w_0 > 0$, and crack area $|C_0| > 0$, let
		\begin{equation}
			\mathcal{A}_\alpha = \left\{ d \in \mathbb{N} : d > w_0 \;\text{ and }\; d \leq 2\sqrt{\frac{\alpha|C_0|}{1 + \varepsilon(d)}} \right\}
		\end{equation}
		be the set of feasible mesh sizes. The \emph{optimal mesh size} is
		\begin{equation}\label{eq:deltamax}
			\Delta_{\max}(\alpha) = \begin{cases} \max \mathcal{A}_\alpha, & \text{if } \mathcal{A}_\alpha \neq \emptyset, \\[3pt] \lceil w_0 \rceil + 1, & \text{if } \mathcal{A}_\alpha = \emptyset. \end{cases}
		\end{equation}
	\end{definition}
	
	The value $\Delta_{\max}(\alpha)$ is found by testing $d = \lceil w_0\rceil + 1, \lceil w_0\rceil + 2, \ldots$ and keeping the largest feasible one. If no candidate is feasible, $\mathcal{A}_\alpha = \emptyset$, the smallest candidate $\lceil w_0\rceil + 1$ is retained; the bound~\eqref{eq:suff_cond} then need not hold, so in this regime $\Delta_{\max}$ should be read as a design value rather than a certified miss-probability bound. This fallback is confined to the strictest tolerances: for instance, at $\alpha = 0.01$ all crack widths in Table~\ref{tab:params} take the value $\lceil w_0\rceil + 1$.

	For the numerical experiments in Section~\ref{ssec:semisynthetic}, we model cracks as rectangles of size $n \times w_0$ on a face of $\widetilde{W} = [1,n]^3$, so that $|C_0| = n w_0$ and
	\begin{equation}\label{eq:eps_rect}
		\varepsilon(d) = \frac{w_0}{d}\log^2\!\left(\frac{w_0}{d}\right) + \frac{d^2}{n w_0}.
	\end{equation}
	The optimal mesh size for this case is then $\Delta_{\max}(\alpha)$ of Definition~\ref{def:deltamax}, evaluated with this $|C_0|$ and $\varepsilon(d)$; in particular the fallback $\lceil w_0\rceil+1$ applies when no integer $d>w_0$ meets the bound.
	
	With $\Delta$ fixed, the graph $G^*$ and its connected components are fully determined. Since these components may arise from cracks, air pores, or noise, a classification criterion is needed to distinguish between them.


	\subsection{Classification via connected components}
	\label{ssec:classification}
	
	Recall that the goal is to localize crack-containing regions within the
	image. To this end, the binary image $L_\mathcal{S}(I)$ is partitioned
	into cubic subimages $\{A_q\}_{q \in \mathcal{Q}}$, where
	$\mathcal{Q} = \{1,\ldots,g_1\} \times \{1,\ldots,g_2\} \times \{1,\ldots,g_3\}$, each of which is classified
	independently as crack-containing or crack-free. Based on the graph
	$G^*_q = (K_q, E_{K_q})$ constructed on the selected face
	$\mathcal{W}_q$, DFS identifies a partition of the vertex set $K_q$
	into connected components $M_q = \{M^1_q, \ldots, M^m_q\}$.
	To distinguish crack-induced components from those arising from noise
	or air pores, we apply two criteria in sequence: a boundary
	intersection filter, which discards interior artefacts, followed by a
	cardinality threshold that identifies crack-induced components by their
	size.
	
	\textbf{Boundary intersection filter.}
	A large crack traversing the subimage $A_q$ must enter and exit through
	the boundary $\partial A^*_q$ of the selected face. Consequently,
	any connected component in $G^*$ induced by a crack must contain
	at least one vertex lying on $\partial A^*_q$. Components arising from air pores or noise, by contrast, remain small and do not form elongated structures spanning the face. All components $M^i_q$
	satisfying $M^i_q \cap \partial A^*_q = \emptyset$ are therefore
	discarded before the cardinality threshold is applied. This step
	removes a large fraction of spurious interior components at
	negligible computational cost, since it requires only a boundary
	check on each vertex.
		
	{\textbf{Cardinality threshold $\tau$.}
		A crack crossing the face $\mathcal{W}$ of a subimage of side $n$
		produces a boundary component of high cardinality in $G^*$, whereas
		components arising from noise or air pores remain small.
		We formalise this by a threshold $\tau$: the subimage $A_q$ is
		classified as containing a crack if
		$$\max_{i=1,\ldots,m} \#M_q^i > \tau.$$
		
		The threshold $\tau$ is intuitively derived from the number of lattice lines
		intersected by a crack traversal, taking both coordinate directions
		into account. Consider a crack of length $n$ oriented at angle
		$\theta \sim \mathrm{Uniform}[0, \pi/2]$ with respect to the
		horizontal axis. The numbers of intersected vertical and horizontal
		lattice lines of $\Delta\mathbb{Z}^2$ are
		$$k_x(\theta)
		= \left\lfloor \frac{n|\cos\theta|}{\Delta} \right\rfloor,
		\qquad
		k_y(\theta)
		= \left\lfloor \frac{n|\sin\theta|}{\Delta} \right\rfloor.$$
		
		If a crack is present in $A_q$, it must intersect the face
		$\mathcal{W}_q$. Depending on its orientation $\theta$, it will
		cut either more vertical or more horizontal lines of
		$\Delta\mathbb{Z}^2$, with the larger count being
		$\max(k_x(\theta), k_y(\theta))$. Each lattice line intersected by the crack has at least one vertex of $K$ within $\ell^\infty$-distance $\Delta$.
		Therefore, if a connected component forms the boundary of a
		crack, its cardinality must satisfy
		$$\#M_q^i
		\;\geq\;
		\max\!\bigl(k_x(\theta),\,k_y(\theta)\bigr)
		\;=\;
		\left\lfloor
		\frac{n\,\max\!\bigl(
			|\cos\theta|,\,|\sin\theta|
			\bigr)}{\Delta}
		\right\rfloor.$$
			
	The expected value of maximum in this lower bound over a uniformly random
	orientation is
	\begin{align}
		\mathbb{E}\!\left[
		\max\!\bigl(|\cos\theta|,|\sin\theta|\bigr)
		\right]
		&= \frac{2}{\pi}
		\int_0^{\pi/2}
		\max(\cos\theta,\sin\theta)
		\,\mathrm{d}\theta \notag \\
		&= \frac{2}{\pi}
		\left[
		\int_0^{\pi/4} \cos\theta\,\mathrm{d}\theta
		+
		\int_{\pi/4}^{\pi/2} \sin\theta\,\mathrm{d}\theta
		\right] \notag \\
		&= \frac{2}{\pi}
		\left[
		\Bigl[\sin\theta\Bigr]_0^{\pi/4}
		+
		\Bigl[-\cos\theta\Bigr]_{\pi/4}^{\pi/2}
		\right] = \frac{2\sqrt{2}}{\pi}.
		\label{eq:Ek} \notag 
	\end{align}
	Therefore,
	$$
	\mathbb{E}\!\left[
	\max\!\bigl(k_x(\theta), k_y(\theta)\bigr)
	\right]
	= \frac{2\sqrt{2}\,n}{\pi\Delta}.
	$$
	This motivates setting
	\begin{equation}\label{eq:tau1}
		\tau
		=
		\left\lfloor \frac{2\sqrt{2}\,n}{\pi\Delta} \right\rfloor.
	\end{equation}
	
	The value \eqref{eq:tau1} follows directly from the geometry: any crack
	spanning the face produces a component of cardinality at least
	$\tau$ on average, regardless of its orientation, while components
	from noise or air pores, which do not traverse the face, are
	expected to remain well below this value. However, in cases where
	the Maximal Hessian Entry filter detects a large air pore or
	aggregate boundary, its boundary component in $G^*$ may attain a
	cardinality comparable to $\tau$, leading to a false positive
	classification. Such misclassifications are an inherent limitation of the
	pre-detection step, which prioritises recall over precision
	by design.
	
	\subsection{Full procedure}
	\label{ssec:algo}
	
	We now summarise the complete crack pre-detection procedure for a 3D
	grayscale CT image $I$.
	
	\begin{enumerate}
		\item \textbf{Binary image computation.}
		Apply the Maximal Hessian Entry filter at scales
		$\mathcal{S} = \{s_1, \ldots, s_r\}$ to obtain the binary image
		$L_\mathcal{S}(I)$.
		
		\item \textbf{Subimage partition.}
		Partition $L_\mathcal{S}(I)$ into cubic subimages
		$A_q = \{L_\mathcal{S}(I,p) : p \in \widetilde{W}_q\}$, where
		$\widetilde{W}_q = [a_q, b_q]^3 \cap \mathbb{Z}^3$
		for $q \in \mathcal{Q}$.
		
		\item \textbf{Face selection.}
		For each $A_q$, select the face $\mathcal{W}_q$ of
		$\partial[a_q,b_q]^3$ with the maximum number of foreground
		pixels and form the 2D slice $A^*_q$.
		
		\item \textbf{Mesh size selection.} Compute $\Delta_{\max}(\alpha)$ from Definition~\ref{def:deltamax} with $|C_0| = n w_0$ and $\varepsilon(d)$ as in~\eqref{eq:eps_rect}.
		
		\item \textbf{Graph construction.}
		For a selected $\Delta \leq \Delta_{\max}(\alpha)$, construct
		the graph $G^*_q = (K_q, E_{K_q})$ on $A^*_q$ following
		Definition~\ref{def:graph}.
		
		\item \textbf{DFS and component extraction.}
		Run DFS on $G^*_q$ to obtain the partition of $K_q$ into
		connected components $M_q = \{M^1_q, \ldots, M^m_q\}$.
		
		\item \textbf{Boundary filter.}
		Discard all components $M^i_q$ with
		$M^i_q \cap \partial A^*_q = \emptyset$.
		
		\item \textbf{Crack classification.}
		Classify $A_q$ as crack-containing if
		$\max_{i=1,\ldots,m} \#M^i_q > \tau$, where $\tau$ is given
		by~\eqref{eq:tau1}.
	\end{enumerate}
	
	The computational complexity of the DFS classification per subimage is
	$O(\#K_q + \#E_{K_q}) \ll O(n^2)$ for $\Delta \gg 1$, which is
	substantially cheaper than operating on the full pixel graph of the
	face. The complexity of the full pipeline is dominated by the Hessian
	filter, which runs in $O(\#W)$ on the entire
	volume and is therefore linear in the number of voxels. The
	efficiency gain of the proposed framework lies in the DFS
	classification step, which adds only negligible overhead on top
	of the filtering stage, as confirmed by the runtimes in
	Table~\ref{tab:runtime}.

	\section{Experimental evaluation}
	\label{sec:experimental-evaluation}
	
	\subsection{Datasets and experimental setting}
	\label{sec:data-acquisition}
	
	The experimental evaluation is conducted on two classes of
	three-dimensional CT data, reflecting both controlled and realistic
	conditions. The semi-synthetic dataset provides voxel-level ground truth
	for quantitative assessment, while the real CT scans allow us to examine
	the behaviour of the method under practical imaging conditions.
	
	\textbf{Semi-synthetic dataset.}
	The semi-synthetic data are drawn from the VoroCrack3D dataset
	\cite{JUNG2024110474}, a publicly available collection of
	$1\,344$ annotated concrete volumes of size $400 \times 400 \times 400$
	voxels. In this dataset, synthetic crack structures are embedded into
	real $\mu$CT scans of crack-free concrete using the minimum-weight surface
	construction of \cite{Jung2023}, yielding realistic crack geometries together
	with exact voxel-level ground truth.
	From this dataset, we select $48$ volumes from the Normal Concrete (NC)
	and High-Performance Concrete (HPC) subsets (cases 4a-9d), covering both
	unbranched and multiscale cracks with widths ranging from $1$ to $7$
	voxels under multiple noise levels. 
	
	\textbf{Real CT scans.}
	To assess performance on real data, we use three concrete images
	acquired with the Gulliver CT system \cite{gulliver2025} at the
	University of Kaiserslautern, with voxel sizes between $50\,\mu$m and
	$300\,\mu$m. In contrast to the semi-synthetic data, no voxel-level
	annotations are available for these scans.

	\subsection{Parameters setting}
	\label{ssec:params}
	
	The proposed method has a single free parameter $\alpha \in (0,1)$, the
	prescribed upper bound on the probability of missing a crack.
	All other algorithmic quantities are derived from $\alpha$, the
	crack width $w_0$, and the subregion side length $n$, fixed
	at $n = 50$ throughout. In the semi-synthetic experiments,
	$w_0$ is known exactly from the dataset construction and is used
	directly as a physical parameter. For real CT scans, where crack
	width $w_0$ varies along the trajectory and is not known in advance, it should be understood as a tuning parameter. 
	
	The choice of $n$ governs the quality of
	the lattice graph $G^*$ on which the DFS operates. If $n$ is too
	small, the face $\mathcal{W}$ contains too few lattice nodes
	$V_\Delta = \mathcal{W} \cap \Delta\mathbb{Z}^2$: even at the
	minimum admissible mesh size $\Delta = \lceil w_0 \rceil + 1$, the
	vertex boundary $K$ is nearly empty, the connected components of
	$G^*$ carry little geometric information, leading to high false negative rate. Conversely, if $n$ is too large, the subregion
	$\widetilde{W}$ covers an excessively large portion of the image,
	and the pre-detection step looses its ability to localise crack-containing
	regions with sufficient spatial precision for subsequent exact
	segmentation methods. The value $n = 50$ balances these two
	requirements: it is large enough that the lattice graph $G^*$
	contains $\lfloor n/\Delta \rfloor^2$ nodes on each face, yielding
	well-connected components whose cardinality reliably distinguishes
	crack-induced boundary paths from pore-induced fragments.
	
	The crack cross-section on a face of $\widetilde{W}$ is modelled as a
	rectangle of area $|C_0| = n w_0$. With $\varepsilon(\Delta)$ and
	$\Delta_{\max}(\alpha)$ defined in~\eqref{eq:eps_shape}
	and~\eqref{eq:deltamax}, the cardinality threshold is
	\begin{equation}
		\tau = \left\lfloor \frac{2\sqrt{2}\,n}{\pi\,\Delta_{\max}} \right\rfloor.
	\end{equation}
	Table~\ref{tab:params} reports the derived pairs $(\Delta_{\max}, \tau)$
	for the four crack widths present in the VoroCrack3D evaluation subset
	and all four tolerance levels considered.
	
	\begin{table}[ht]
		\centering
		\small
		\caption{Derived parameters $(\Delta_{\max},\,\tau)$ per crack width $w_0$
			and tolerance $\alpha$, with subregion side $n = 50$ and
			$|C_0| = n\,w_0$.}
		\label{tab:params}
		\setlength{\tabcolsep}{4pt}
		\renewcommand{\arraystretch}{1.3}
		\begin{tabular}{lc *{4}{r@{,\ }l}}
			\hline\hline
			$w_0$ & $|C_0|$
			& \multicolumn{2}{c}{$\alpha = 0.01$}
			& \multicolumn{2}{c}{$\alpha = 0.05$}
			& \multicolumn{2}{c}{$\alpha = 0.10$}
			& \multicolumn{2}{c}{$\alpha = 0.20$} \\
			\hline
			$1$ & $\phantom{0}50$
			& $\Delta_{\max}{=}2$ & $\tau{=}22$
			& $\Delta_{\max}{=}2$ & $\tau{=}22$
			& $\Delta_{\max}{=}3$ & $\tau{=}15$
			& $\Delta_{\max}{=}4$ & $\tau{=}11$ \\
			$3$ & $150$
			& $\Delta_{\max}{=}4$ & $\tau{=}11$
			& $\Delta_{\max}{=}4$ & $\tau{=}11$
			& $\Delta_{\max}{=}6$ & $\tau{=}7$
			& $\Delta_{\max}{=}8$ & $\tau{=}5$  \\
			$5$ & $250$
			& $\Delta_{\max}{=}6$ & $\tau{=}7$
			& $\Delta_{\max}{=}6$ & $\tau{=}7$
			& $\Delta_{\max}{=}8$ & $\tau{=}5$
			& $\Delta_{\max}{=}10$ & $\tau{=}4$ \\
			$7$ & $350$
			& $\Delta_{\max}{=}8$ & $\tau{=}5$
			& $\Delta_{\max}{=}8$ & $\tau{=}5$
			& $\Delta_{\max}{=}10$ & $\tau{=}4$
			& $\Delta_{\max}{=}12$ & $\tau{=}3$ \\
			\hline\hline
		\end{tabular}
	\end{table}
	
	\subsection{Semi-synthetic evaluation}
	\label{ssec:semisynthetic}
	
	We evaluate the proposed method on the 48 semi-synthetic volumes described
	in Section~\ref{sec:data-acquisition}. Each volume belongs to either the
	Normal Concrete (NC) or the High-Performance Concrete (HPC) subset of
	VoroCrack3D \cite{JUNG2024110474} and carries voxel-level ground truth
	annotations. Since the method operates at the subregion level, classifying
	each cubic block $A_q$ as crack-containing or crack-free, we report
	standard binary classification metrics computed over subregions: Recall~$R$,
	Precision~$P$, F1 score $= 2PR/(P+R)$, and Intersection over Union
	$\mathrm{IoU} = |D \cap G| / |D \cup G|$, where $D$ and $G$ denote the
	sets of subregions classified as crack-containing by our method and by the
	ground truth, respectively. A subregion is treated as a single unit regardless of the number of crack voxels it contains, so the evaluation is performed at the subregion level and is therefore coarser than voxel-level measures. This is consistent with the goal of localising crack-containing regions rather than producing an exact voxel-wise segmentation.
	
	Figure~\ref{fig:qualitative_nc} provides representative examples from the
	NC subset while $\alpha=0.05$ and subregion side length
	$n=50$. For each selected image, the top row shows the input CT slice with
	the subregion-level classification overlay and the ground truth crack
	boundaries, the middle row shows the corresponding Hessian-based binary image
	with the subregion partition and component sizes, and the bottom row shows
	the graph $G^*$ on the selected face of several representative crack-containing
	subcubes.
	
	\begin{figure}[ht]
		\centering
		\includegraphics[width=\textwidth]{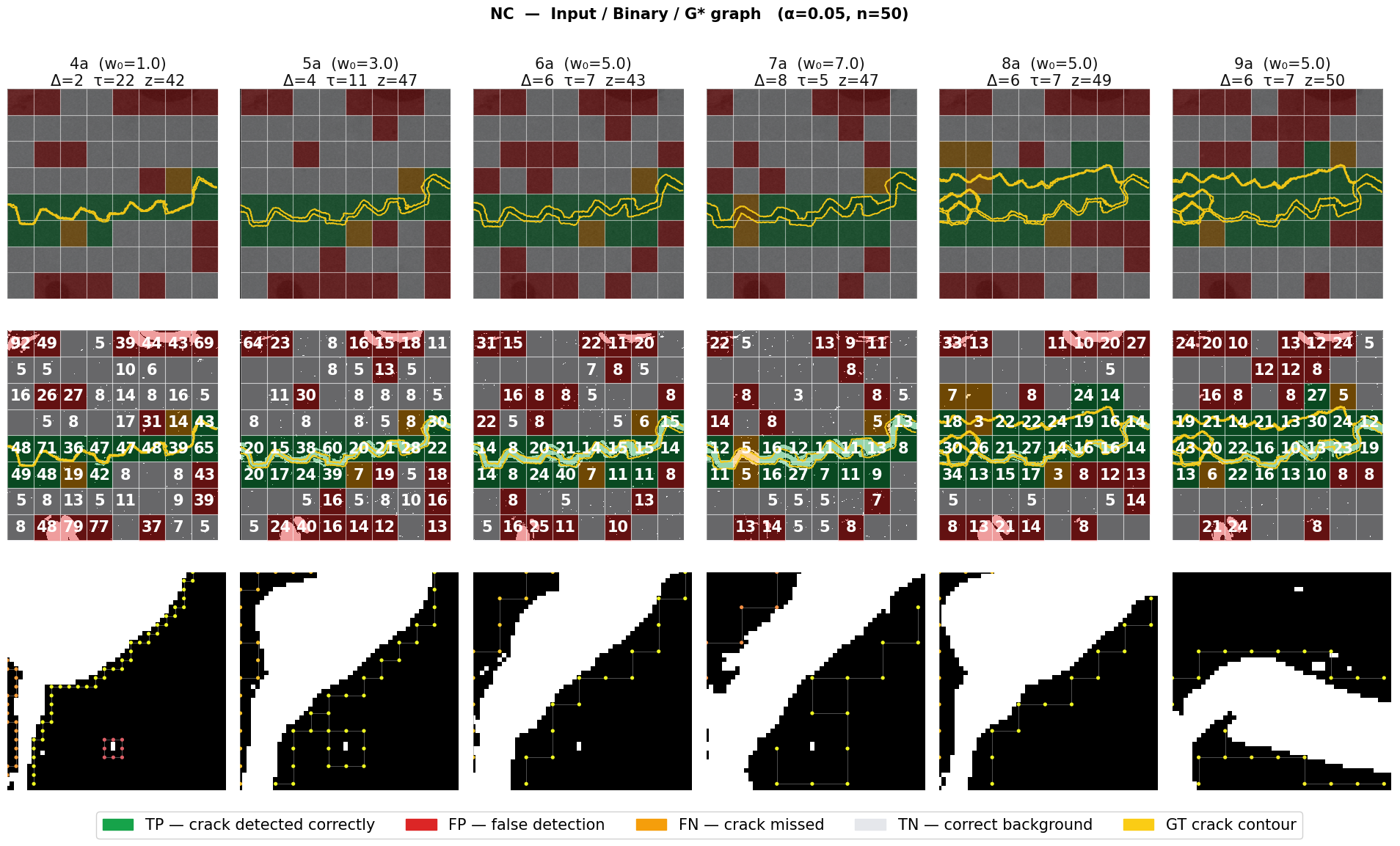}
		\caption{Representative examples from the NC subset at $\alpha=0.05$
			and subregion side length $n=50$.}
		\label{fig:qualitative_nc}
	\end{figure}
	
	Figure~\ref{fig:syntheticmetric} reports Recall, Precision, F1, and IoU
	per image for both subsets and all four tolerance levels
	$\alpha \in \{0.01,\,0.05,\,0.10,\,0.20\}$.
	For the NC subset, recall remains high across the full image sequence,
	typically between approximately $0.85$ and $0.97$. For the HPC subset,
	recall is lower and more variable, with several images falling to about
	$0.73$-$0.77$. In both material classes, the recall curves corresponding
	to the four values of $\alpha$ are nearly indistinguishable, indicating
	that the empirical recall is only weakly affected by $\alpha$ within the
	tested range.
	\begin{figure}[ht]
		\centering
		\includegraphics[width=\textwidth]{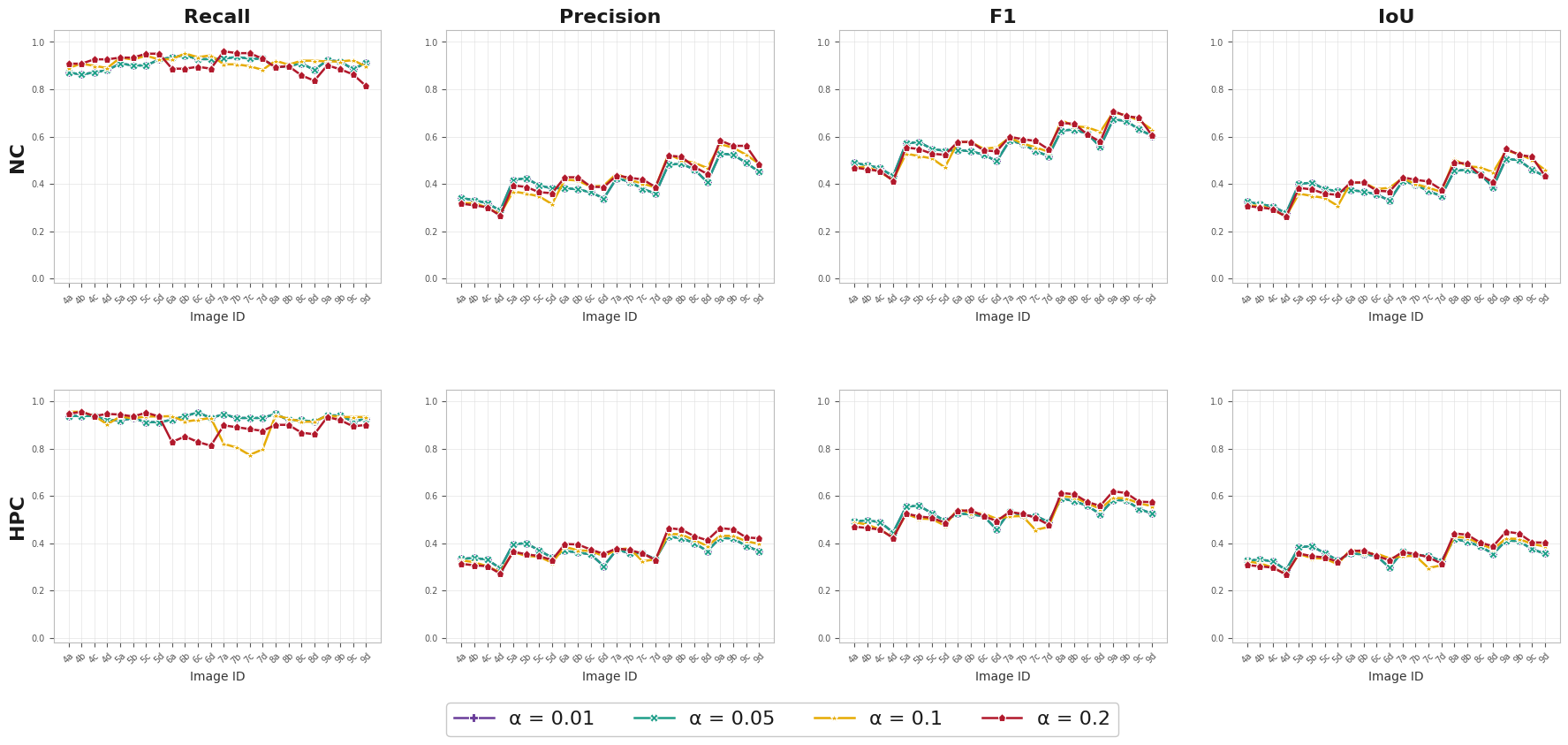}
		\caption{Performance metrics (Recall, Precision, F1, IoU) per image
			for the NC (top row) and HPC (bottom row) subsets of VoroCrack3D,
			across four tolerance values $\alpha$.
			Images are ordered by crack-width group:
			4a-4d ($w_0=1$), 5a-5d ($w_0=3$), 6a-6d ($w_0=5$),
			7a-7d ($w_0=7$), 8a-9d (multiscale).}
		\label{fig:syntheticmetric}
	\end{figure}
	
	Precision, F1, and IoU show a common pattern in both subsets. Across the
	ordered image sequence, these metrics display an overall upward trend from
	the narrow-crack groups toward the wider and multiscale groups. 
	In addition, a recurrent oscillation is visible within each group of four
	images. This is consistent with the construction of VoroCrack3D, where the
	suffixes a-d correspond to increasing added-noise levels: a denotes no
	added noise, while b, c, and d are obtained by adding uniform noise with
	progressively larger amplitude.

	Overall, the semi-synthetic experiments show that the proposed method is
	effective as a subregion-level pre-detection stage. Its performance is
	consistently stronger on NC than on HPC, and its behaviour is stable across
	the tested values of $\alpha$. Since $\alpha = 0.05$ performs comparably to
	the other tested values while allowing a slightly coarser lattice than
	$\alpha = 0.01$, we adopt it as the default choice in the remaining
	experiments.

	\subsection{Real CT scan evaluation}
	\label{ssec:real}
	
	We now evaluate the proposed method on three real CT scans, namely HPC3,
	NC1, and NC12. The resolution is $850 \times 400 \times 1050$ voxels for
	HPC3 and $900 \times 900 \times 1050$ voxels for both NC1 and NC12. In contrast to the semi-synthetic dataset, no voxel-level
	ground truth crack annotations are available for these data. As a visual reference, we use the
	crack contour obtained from a U-Net-based segmentation, which is overlaid
	in yellow in Figure~\ref{fig:real2}. At the subregion level, a cube $A_q$
	is treated as reference-positive whenever the reference contour intersects
	the corresponding subregion $\widetilde{W}_q$.
		
	\textbf{Parameter setting.}
	Unlike the semi-synthetic data, the real CT scans do not admit a single
	well-defined crack width. The observed crack thickness varies along the
	trajectory, so $w_0$ is chosen here as a representative scale for parameter
	selection. For HPC3, we take $w_0 = 5$ to reflect the broader visible crack band, leading to $(\Delta,\tau) = (6,7)$. For NC1 and NC12, we use the smaller value $w_0 = 3$, leading to $(\Delta,\tau) = (4,11)$, in order to retain a finer lattice, reduce the risk of missing narrower crack sections, and assess the effect of $w_0$ on the resulting detection. Thus, in the real-image experiments, $w_0$ should be understood
	as a practical tuning parameter for multiscale crack localization rather
	than as an exact measured width. All three scans are processed with the
	default setting $\alpha = 0.05$ and subregion side length $n = 50$.
	
	\begin{figure}[htbp]
		\centering
		\includegraphics[width=0.9\textwidth]{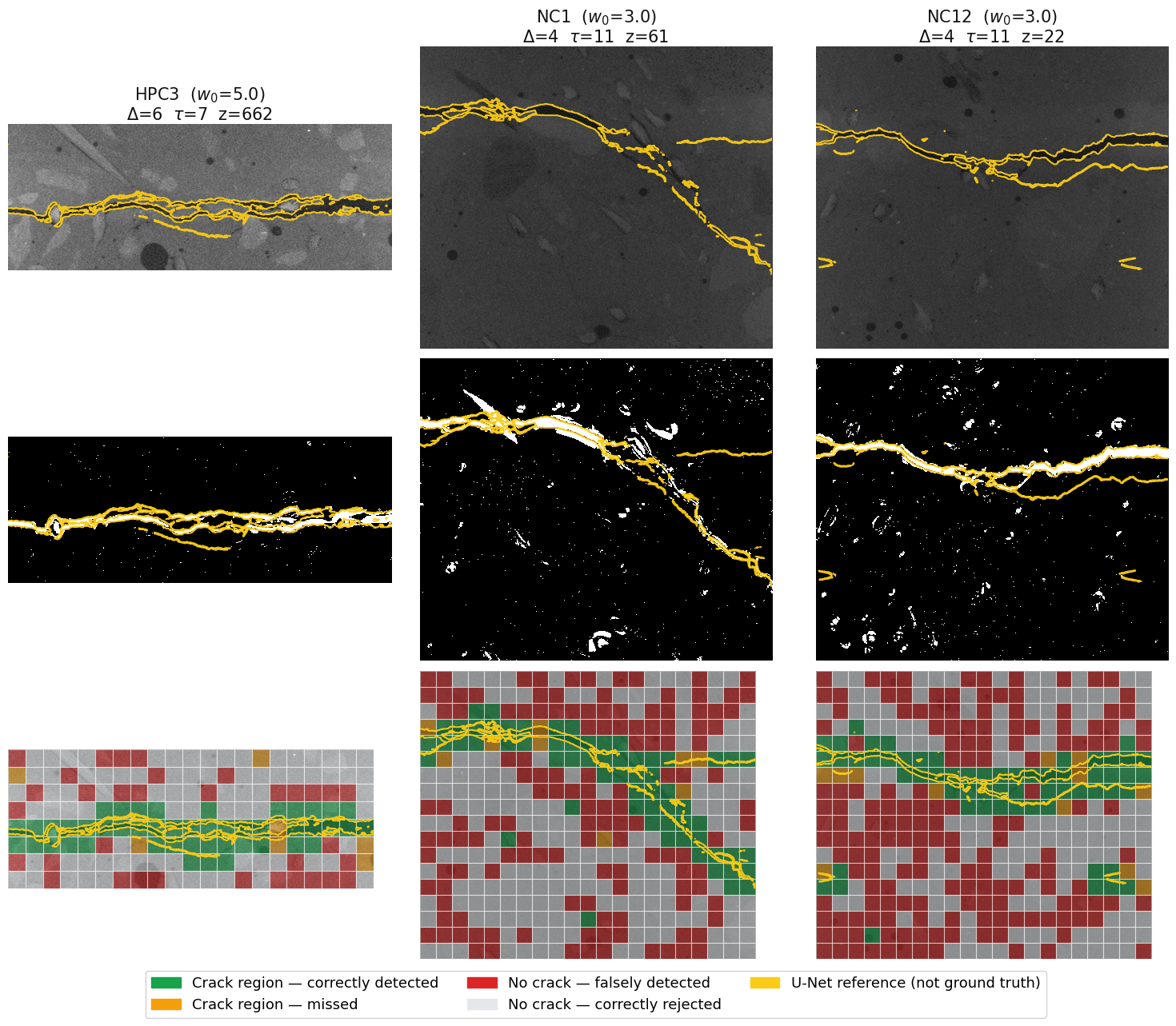}
		\caption{Representative real CT scan results at $\alpha = 0.05$ and
			subregion side length $n = 50$. From left to right: HPC3, NC1,
			and NC12. For each specimen, the top row shows the input CT slice
			with the reference crack contour, the middle row shows the
			Hessian-based binary image $L_{\mathcal{S}}(I)$ with the same
			contour, and the bottom row shows the final DFS-based subregion classification, compared with the U-Net segmentation. Green: true positive; red: false positive;
			orange: false negative; light gray: true negative; yellow:
			reference crack contour.}
		\label{fig:real2}
	\end{figure}
	
	\textbf{Qualitative analysis.} Several common features are visible across all three scans. First, the
	Maximal Hessian Entry filter recovers the main crack structure in the
	binarized image, but also produces a substantial number of additional
	foreground regions. These spurious responses are caused by pores, aggregate
	boundaries, and other structural inhomogeneities that are typical of real CT
	data. Second, the subsequent DFS-based classification concentrates the
	detections around the dominant elongated crack path, thereby suppressing a large portion of non-crack structures. Thus, even on real scans, the proposed method serves as an effective pre-localization step that substantially reduces the search space compared with the raw binary image.
	
	However, the bottom row of Figure~\ref{fig:real2} shows that the
	final detection still contains a substantial number of false positive
	subregions. This is consistent with the behaviour of the Hessian-based
	binary image in the middle row, where many non-crack structures are also
	enhanced. The DFS step reduces non-crack foreground while preserving the dominant crack structure, but false positives remain. Thus, the method serves as a pre-detection and localization step rather than an exact segmentation method.

	Overall, the real-scan examples indicate that the method generalizes to
	real data in the sense that it consistently highlights the main
	crack-containing region. While the false positive rate remains substantial,
	the method rarely misses actual crack subregions, which is the critical
	property for a pre-detection stage. The detected subregions can therefore
	serve as a focused search space for more precise subsequent analysis:
	by restricting the computation of additional geometric statistics
	\cite{vitalii24} to the detected subregions only, the overall
	pipeline avoids processing the vast crack-free majority of the volume.

	\section{Discussion}
	\label{sec:discussion}

	\subsection{Sensitivity of the method and the role of \texorpdfstring{$\alpha$}{alpha}}
	\label{ssec:discussion-sensitivity}
	
	The experiments show that the method is only weakly affected by $\alpha$
	within the tested range. In Figure~\ref{fig:syntheticmetric}, the Recall
	curves for $\alpha \in \{0.01,0.05,0.10,0.20\}$ are nearly
	indistinguishable. This is consistent with the theory in
	Section~\ref{ssec:probability}: $\alpha$ controls the lattice-miss event in
	the DFS sampling step, not the total empirical error of the full pipeline.
	It should be noted that the empirical recall reflects the miss rate over all crack-containing subregions, whereas $\alpha$ controls only the idealized lattice-miss bound used for mesh-size selection. Since most missed detections arise from the Hessian binarization stage rather than the lattice sampling, changes in $\alpha$ have little effect on the overall recall.
	
	In practice, the final sensitivity depends on several coupled stages. The
	Hessian-based binarization already determines which candidate crack voxels
	remain visible, as seen in Figures~\ref{fig:hessian1} and
	\ref{fig:hessian2}. The reduction to a single face
	(Figure~\ref{fig:facet2d}) may weaken the crack signature, and the final
	decision still depends on whether the connected component in $G^*$ exceeds
	the threshold $\tau$ (Figure~\ref{fig:graph}). Thus, the effective
	sensitivity is controlled by the full pipeline rather than by $\alpha$
	alone.
	
	The same conclusion is supported by Figure~\ref{fig:real2}. For real CT scans, $w_0$ enters through the choice of $\Delta$ and $\tau$ and represents only an approximate crack scale. Consequently, the sensitivity is governed jointly by $w_0$, $\Delta$, and $\tau$. The results indicate that the method is robust to moderate changes in $\alpha$.
	
	\subsection{Main limitations of the pre-detection approach}
	\label{ssec:discussion-limitations}
	
	The main limitation of the method arises from the Hessian-based binarization stage. Figures~\ref{fig:hessian1} and \ref{fig:hessian2} show that the Maximal Hessian Entry filter preserves the crack band but also enhances pores and other non-crack structures. As a result, false positive subregions remain visible around the crack path even after the DFS-based step. A natural refinement is therefore to compute additional geometric statistics inside the detected subregions, as in \cite{vitalii24}.
	
	A second limitation lies in the face selection step. While reducing each cube to a single most informative face provides the main computational gain, it may introduce false negatives if the crack intersects the selected boundary only weakly (Figure~\ref{fig:facet2d}).
	
	A third limitation concerns the geometric assumptions underlying the theory. The miss-probability analysis in Section~\ref{ssec:probability} is based on an idealized planar convex crack cross-section, whereas real cracks and the semi-synthetic HPC cases are often more irregular. This discrepancy is reflected in Figures~\ref{fig:syntheticmetric} and \ref{fig:real2}, where the crack is detected but the localization includes both false positives and missed regions.
		
	\subsection{Run time}
	\label{ssec:runtime}
	
	Table~\ref{tab:runtime} reports the runtimes for the three real
	CT scans at $\alpha=0.05$ and subregion side length $n=50$. The Hessian
	filter was executed on GPU (NVIDIA RTX 3090) at the four scales
	$\mathcal{S}=\{1,3,5,10\}$, while the DFS-based classification was executed on CPU.
	
\begin{table}[ht]
	\centering
	\small
	\caption{Runtimes for the three real CT scans
		($\alpha=0.05$, $n=50$). The Hessian filter runs on GPU
		(NVIDIA RTX 3090, $\mathcal{S}=\{1,3,5,10\}$, chunk size 64);
		the DFS classification runs on CPU.}
	\label{tab:runtime}
	\setlength{\tabcolsep}{5pt}
	\renewcommand{\arraystretch}{1.3}
	\begin{tabular}{lccccc}
		\hline\hline
		Scan & Shape & $(\Delta,\tau)$
		& Filtering (s) & DFS (s) & Subcubes \\
		\hline
		HPC3  & $850\times400\times1050$  & $(6,7)$
		& 58.1 & 1.5 & 2\,856 \\
		NC1   & $900\times900\times1050$  & $(4,11)$
		& 142.1 & 6.3 & 6\,804 \\
		NC12  & $900\times900\times1050$  & $(4,11)$
		& 142.3 & 6.1 & 6\,804 \\
		\hline\hline
	\end{tabular}
\end{table}
	
	The runtime is dominated by the Hessian-based binarization stage. For all
	three scans, the DFS-based subregion classification is much faster, taking
	only $1.5$-$6.3$ seconds compared with $58.1$-$142.3$ seconds for the
	Hessian filter. This is consistent with the structure of the method: the
	Hessian filter is applied to the full three-dimensional image, whereas DFS
	operates only on the sparse graph $G^*$ built on selected subregions.
	
	Overall, the results confirm that the computational cost of the proposed
	framework lies mainly in the multiscale filtering step, while the DFS-based
	localization itself is inexpensive and therefore suitable as a screening
	stage before more exact downstream segmentation.

	\section{Conclusions}
	\label{sec:conclusion}
	This paper proposed a fast pre-detection framework for crack localization in
	large three-dimensional CT images of concrete. The method combines the
	Maximal Hessian Entry filter with a DFS-based analysis of connected
	components on selected subregion faces. Its main purpose is not exact
	voxel-wise crack segmentation, but the efficient identification of
	crack-containing regions that can subsequently be analysed in more detail by
	slower segmentation procedures.
	
	The experimental results on both semi-synthetic and real CT data show that the proposed method preserves the dominant crack structure and reliably localizes crack-containing areas at the subregion level. On the VoroCrack3D data, the method achieves consistently good performance, with higher recall for the NC subset than for the more challenging HPC cases. On real CT scans, visual comparison indicates that the main crack trajectory is successfully identified, despite the presence of false positives.
	
	However, the precision remains limited due to high false positives, so the method should be interpreted as a filtering stage that reduces, rather than fully eliminates, non-crack regions.
	
	From the computational point of view, the main cost of the framework lies in
	the multiscale Hessian filtering step, whereas the DFS-based localization is
	comparatively inexpensive. This makes the method attractive as a screening
	stage in large-scale CT workflows, where a full high-resolution analysis of
	the entire image would be prohibitively expensive. The proposed approach balances sensitivity and computational efficiency while retaining geometric interpretability.
	
	From the theoretical point of view, the paper also provides an explicit
	probabilistic criterion for selecting the mesh size through an upper bound
	on the lattice-miss event for an idealized crack cross-section. Although the
	full empirical behavior depends on the entire pipeline and on the geometry
	of real cracks, this analysis gives a principled basis for parameter
	selection.
	
	Overall, the proposed method is a pre-detection and localization tool for large CT images of concrete. It is computationally efficient, as the DFS-based step operates on a sparse graph, while the main cost is confined to the Hessian filtering stage. Despite high false positives arising from non-crack structures, the method reliably preserves the dominant crack structure and achieves high sensitivity at the subregion level. As a result, it provides an effective screening stage that substantially reduces the search space for subsequent, more precise analysis. Future work may address adaptive parameter selection for real multiscale cracks, improved suppression of false positives through additional geometric statistics, and integration with more accurate statistical or learning-based segmentation methods. Furthermore, it can accelerate the training of neural networks by removing homogeneous background regions before the learning step.
	
\section*{Acknowledgements} Vsevolod Chernyshev would like to thank Anton Ayzenberg for the
fruitful discussion.
\section*{Funding}
This research was funded by the Federal Ministry of Research, Technology and Space of Germany (BMFTR) [grant number 05M20VUA (DAnoBi)] and partially supported by the International Research Center ”Innovation Transportation and Production Systems” of the I-SITE CAP 20-25.
\section*{Declaration of Competing Interest}
The authors declare that they have no known competing financial interests or personal relationships that could have appeared to influence the work reported in this paper.
\section*{Data Availability}

The semi-synthetic CT data used in this study are from the publicly available VoroCrack3D dataset. The real CT scans used in the experimental evaluation are not publicly available because they originate from third-party experimental acquisitions and permission for public redistribution has not been granted.

\bibliographystyle{elsarticle-num}
\bibliography{references}

\end{document}